\documentclass{ijuc}
\usepackage{graphicx}
\usepackage{amsmath, amssymb}
\usepackage{multirow}
\usepackage{multicol}
\usepackage{tabularx}
\usepackage{siunitx}
\usepackage[space]{cite}
\usepackage{adjustbox}
\usepackage{pgfplotstable}
\usepackage[inline]{trackchanges}
\addeditor{CT}
\addeditor{DT}

\begin{document}

\title{Memcapacitive Devices in Logic and Crossbar Applications}

\author{
Dat Tran, S. J.    \email{datran@pdx.edu}
\and 
Christof Teuscher
}

\institute{
Department of Electrical and Computer Engineering\\
Portland State University, Portland, OR, USA
}

\def\received{Received 25 December 2016; In final form 1 April 2017}

\maketitle

\begin{abstract}
Over the last decade, memristive devices have been widely ad\-opted in computing for various conventional and unconventional applications. While the integration density, memory property, and nonlinear characteristics have many benefits, reducing the energy consumption is limited by the resistive nature of the devices. Memcapacitors would address that limitation while still having all the benefits of memristors. Recent work has shown that with adjusted parameters during the fabrication process, a metal-oxide device can indeed exhibit a memcapacitive behavior. We introduce novel memcapacitive logic gates and memcapacitive crossbar classifiers as a proof of concept that such applications can outperform memristor-based architectures. The results illustrate that, compared to memristive logic gates, our memcapacitive gates consume about $7\times$ less power. The memcapacitive crossbar classifier achieves similar classification performance but reduces the power consumption by a factor of about $1,500\times$ for the MNIST dataset and a factor of about $1,000\times$ for the CIFAR-10 dataset compared to a memristive crossbar. Our simulation results demonstrate that memcapacitive devices have great potential for both Boolean logic and analog low-power applications.
\end{abstract}


\section{Introduction}

The ever-growing demand for more speed and lower power in circuit design poses significant challenges for the continuing scaling of today's CMOS technology. Fundamental physical as well as architectural limits lead to new bottlenecks. While the advent of multicore architectures alleviated some of the challenges, more cores do not always mean better: only a fraction of the cores typically operate at full speed because of Amdal's law and power constraints \cite{Esmaeilzadeh2013}. Finding alternative devices and architectures beyond CMOS, beyond Boolean logic, and beyond von Neumann architectures has been a major driver of the unconventional computing community.

Memristive devices \cite{Strukov2008} have been widely adopted in previous years for various conventional and unconventional applications. They have shown great promise for high integration densities as well as low energy consumption \cite{Querlioz2012, Serrano2013, Indiveri2013}, for example for neuromorphic applications \cite{Kim2015,Hu2014,Kim2014,Zhou2014,Taha2014} and for memristor-based logic circuit design \cite{Vourkas2016emerging}. However, the energy consumption of memristors is bounded by the resistive nature of these devices. That is where memcapacitors \cite{Mohamed2015, Biolek2013}, another mem-element, may have further benefits.

Recent work demonstrated a memcapacitive response in a $MoS_2$ mono-layer metal insulator devices  \cite{Khan2016}, in a metal-insulator composite of $Si_3N_4$, $p-Si$, and $BiFeO_3$ \cite{You2016}, in organic polymer layers embedded with graphene sheets \cite{Park2016}, in a nano device of polyvinyl alcohol/cadmium sulphide \cite{Sarma2016}, and in a hafnium oxide ($HfO_x$) on n-type Si substrate \cite{Yang2016}. Mohamed \textit{et al.} discovered that it is possible to construct a memcapacitive device from a memristive metal-oxide composite by adjusting the physical device parameters \cite{Mohamed2015}. The memcapacitive characteristics of the device solely depend on a behavior shape factor (BSF), which is controllable during the fabrication process. Mohamed \textit{et al.} derived a mathematical model that describes the response of a metal-oxide device based on the device state, the capacitive current, and the tunneling current. When the behavior shape factor is less than $0.1$, the capacitive current becomes dominant and the device operates as a memcapacitor \cite{Mohamed2015}. Biolek \textit{et al.} designed a SPICE model that describes the correlation between electrical charge $q$ and voltage $V_C$ using a dependent voltage-controlled current source \cite{Biolek2013}. Their SPICE model produced the predicted results of a bipolar memcapacitive model with threshold through simulations in PSpice, LTspice, and HSPICE. 

Several applications of memcapacitive devices have been proposed, such as the dynamic configurations of transmission lines \cite{Pershin2015}, improving a cellular neural network's density \cite{Yi2015}, a memcapacitive synapse with integrate-and-fire neurons \cite{Pershin2014}, dynamic computing random access memory \cite{Traversa2014}, and biomimetic sensors \cite{Chen2014}. Logic applications, combined with CMOS inverters, have been demonstrated for both memristors \cite{Kvatinsky2012, Abdoli2014} and memcapacitors \cite{Traversa2014}. Similar to memristive logic gates, which can improve the chip density by a factor of $2$ compared to CMOS gates \cite{Cho2015}, memcapacitive logic gates are equally promising for an increased area density. While memristive crossbars are widely adopted for machine learning applications, such as pattern classification \cite{Alibart2013, Zamanidoost2015}, high-speed image processing \cite{Hu2012}, and random access memory \cite{Vourkas2016}, memcapacitive crossbars, to the best of our knowledge, were only introduced in \cite{Strukov2013capacitive, GE2016Memcapacitive} but not fully explored in this context. 

In this paper, we propose two novel memcapacitor applications: (1) binary switching in digital logic and (2) analog computing in a crossbar classifier. Our main contributions include a new set of memcapacitive logic gates as well as a memcapacitor-based crossbar classifier. Our results show that both memcapacitor architectures are significantly more energy-efficient while performing similarly compared to memristor-based architectures. The work expands the foundations of computing with memcapacitive devices and is relevant for applications where low power is critical, such as mobile platforms, the Internet of Things (IoT), and embedded systems.

\section{Background}
Although memcapacitive behaviors were observed in several composite devices \cite{Khan2016, You2016, Park2016, Sarma2016, Yang2016}, only two models are currently available in the literature: the \textit{Biolek} model \cite{Biolek2013} and the \textit{Mohamed} model \cite{Mohamed2015}. These two models are selected for our studies.

The \textit{Biolek} model describes a memcapacitive behavior of an ideal device with a threshold. The memcapacitance $C$ functions as an internal variable $\rho$ and is related to the electric charge $q$ and the applied voltage $V_C$ \cite{Biolek2013}:
\begin{subequations}
 \begin{align*}
     q(t)          &= CV_C,\\
     \frac{dC}{dt} &= f\left(V_C\right)W\left(C, V_C\right),
 \end{align*}
\end{subequations}
where $f()$ is a function that describes the threshold property and $W()$ is a window function. These functions are defined as:
\begin{subequations}
 \begin{align*}
     f\left(V_C\right) &= \beta \left(V_C - 0.5\left[\left|V_C + V_{th}\right| - \left|V_C - V_{th}\right|\right]\right),\\
     W\left(C, V_C\right) &= \theta\left(V_C\right)*\theta\left(C_{high} - C\right) + \theta\left(-V_C\right)* \theta\left(C - C_{low}\right)
 \end{align*}
\end{subequations}

$\beta$ is a device constant expressing how the memcapacitance $C$ changes when $\left|V_C\right| > V_{th}$, $V_{th}$ is a threshold voltage, $\theta()$ is a step function, and $C_{low}$ and $C_{high}$ are the minimum and maximum values of the device's capacitance.

\begin{figure}[!htb]
    \begin{center}
    {\includegraphics [width=\textwidth]{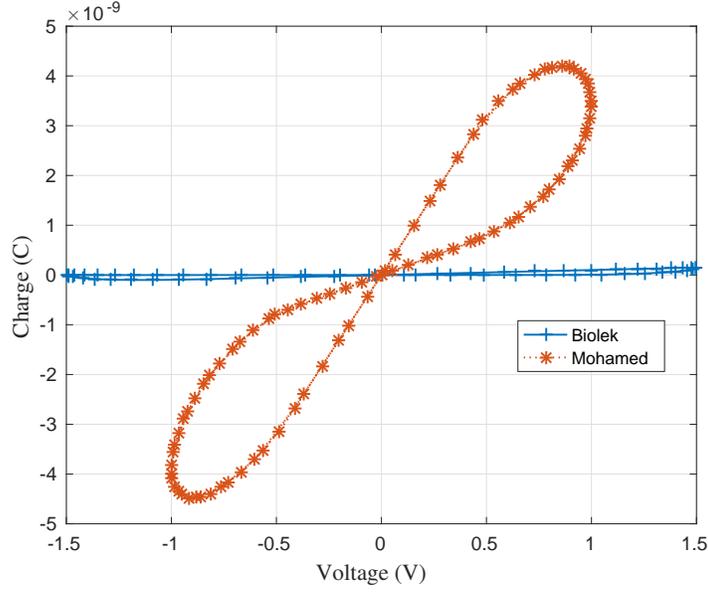}}
    \end{center}
    \caption{Charge-voltage response of the \textit{Biolek} model \cite{Biolek2013} and the \textit{Mohamed} model \cite{Mohamed2015}.}
    \label{fig:QVResponse}
\end{figure}

The \textit{Mohamed} model depicts the memcapacitive response of a metal-dioxide device. The correlations of the device states ($x$ and $m$), the memcapacitance $C$, and applied voltage $v$ are as following \cite{Mohamed2015}:
\begin{subequations}
 \begin{align*}
     q(t)    &= C(x,m,v,t)v(t),\\
     \dot{x} &= \frac{dx}{dt} = f\left(x,v,t\right),\\
     \dot{m} &= \frac{dm}{dt} = f\left(m,v,t\right),
 \end{align*}
\end{subequations}
where $x$ is the filament growth due to ion migrations between the metal-dioxide gap, $m$ is the cross section area of the filament, and $f()$ is a window function defined in \cite{Biolek2009spice}. The memcapacitance $C$ is a function of the device's total capacitance. This function depends on the permittivity of the gap insulator $\varepsilon$, the gap cross section $A$, and the maximum gap thickness $d$. The derivatives of the state variables $x$ and $m$ model the growth/shrinkage of the filament, which is controlled by the tunneling current $i_t(t)$ and the capacitive current $i_c(t)$ \cite{Mohamed2015}.

Fig. \ref{fig:QVResponse} shows the charge-voltage responses of the \textit{Biolek} \cite{Biolek2013} model and the \textit{Mohamed} model \cite{Mohamed2015}. As one can see, the responses follow a pinched hysteresis loop, which is the fundamental characteristic of a mem-device. The threshold voltage of the \textit{Biolek} model was set to 0.8V. The threshold voltage was added to the original \textit{Mohamed} model and the constants were modified to deal with a low input frequency of 1Hz: $K_G = 0.4775, K_S = 0.64, B_G = 2.2475, B_S = 2.75, x_{min} = 0.4, x_{max} = 0.9, m_{min} = 0.01, m_{max} = 0.9, d_1 = \num{5e-10}$, and $d_2 = \num{5e-10}$.

\section{Proposed Memcapacitive Circuits}
\subsection{Memcapacitive Logic Gates}
Logic gates form the fundamental building blocks of digital circuits and architectures. It was proven that both memristors \cite{Kvatinsky2014memristor} and memcapacitive devices \cite{Pershin2015memcomputing} are capable of performing logic operations using material implications. Several studies have shown that logic gates can be realized with memristors \cite{Cho2015, Abdoli2014} and that such gates consume less power and allow for higher integration densities than CMOS gates. The first design of memristive gates was developed for fuzzy logic \cite{Klimo2011memristors}, which was extended to include sorting networks \cite{Nielen2016memristive}. It was show to be compatible with CMOS AND/OR functionality \cite{Kvatinsky2012}. The main idea for designing a memristive logic gate is based on voltage division: several resistors connected in series can scale an applied voltage to different voltages according to their resistance values. Unlike traditional fixed-value resistors, memristors have the ability to alter their resistance to an ON state (low resistance) or OFF state (high resistance). The voltages across them can therefore change dynamically. If the ON/OFF resistance ratio is sufficiently large, each memristor in a memristive gate can operate as a binary switch, analogous to a CMOS switch. Here, we apply the same concept to memcapacitive gates since such devices, when connected in series, can also scale voltages according to their dynamic capacitance.

\begin{figure}[!htb]
    \begin{center}
    {\includegraphics[width=0.8\textwidth]{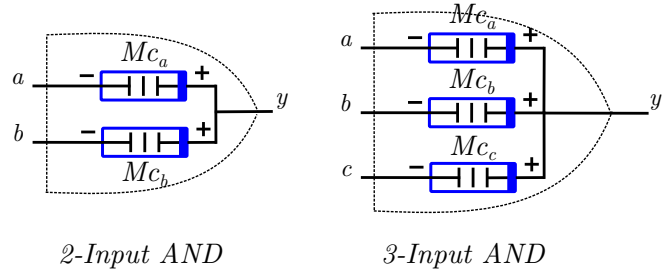}}
    \end{center}
    \caption{2-input and 3-input memcapacitive AND gates.}
    \label{fig:MemcapAND}
\end{figure}

\begin{figure}[!htb]
    \begin{center}
    {\includegraphics[width=0.8\textwidth]{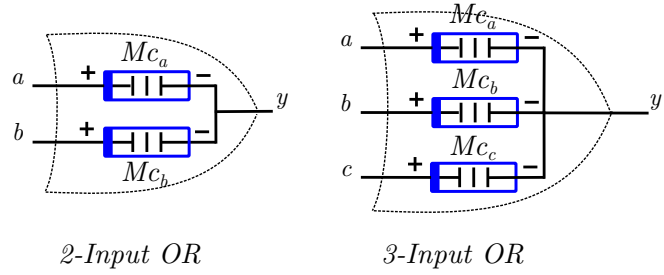}}
    \end{center}
    \caption{2-input and 3-input memcapacitive OR gates.}
    \label{fig:MemcapOR}
\end{figure}

Considering the 2-input gate in Fig. \ref{fig:MemcapAND}, $Mc_a$ and $Mc_b$ are in series with respect to inputs \textit{a} and \textit{b}. The electric charge is the same for both devices:

\begin{align}
    Q_{MCa} &= Q_{MCb}, \nonumber \\
    \left(V_a - V_y\right)*Mc_a  &= \left(V_y - V_b\right)*Mc_b, \nonumber \\
    \left(Mc_a + Mc_b\right)*V_y &= V_a * Mc_a + V_b * Mc_b, \nonumber \\
    V_y &= \frac{V_a * Mc_a + V_b * Mc_b}{Mc_a + Mc_b}. \label{eq:GateOutput}
\end{align}

Assuming that $C_{max} >> C_{min}$ with 0V for logic 0 and 1V for logic 1, we consider four cases for the output $V_y$ according to Eq. \ref{eq:GateOutput}:
\begin{itemize}
  \item{$Va = 0V, V_b = 0V$: $V_y = 0V$}
  \item{$Va = 0V, V_b = 1V$: with their connection polarities, $Mc_a$ is switched to $C_{max}$, $Mc_b$ is switched to $C_{min}$, and the output voltage is:}
    \begin{equation*}
      V_y = \frac{C_{min}}{C_{max} + C_{min}} * V_b < V_{LH} \approx logic \ 0, 
    \end{equation*}
   where $V_{LH}$ is the upper limit voltage for logic 0.
   \item{$Va = 1V, V_b = 0V$: $Mc_a$ is switched to $C_{min}$, $Mc_b$ is switched to $C_{max}$, and the output voltage is:}
    \begin{equation*}
      V_y = \frac{C_{min}}{C_{min} + C_{max}} * V_a < V_{LH} \approx logic \ 0 
    \end{equation*}
  \item{$Va = 1V, V_b = 1V$: the output voltage is:}
    \begin{equation*}
      V_y = \frac{Mc_a + Mc_b}{Mc_a + Mc_b} = 1V
    \end{equation*}
\end{itemize}

The input combinations of $a$ and $b$ along with the output values of $y$ constitute the truth table of an AND gate.

Similarly, for the 2-input OR gate (Fig. \ref{fig:MemcapOR}) and from on the Eq. \ref{eq:GateOutput}, we consider four cases:
\begin{itemize}
  \item{$Va = 0V, V_b = 0V$: $V_y = 0V$}
  \item{$Va = 0V, V_b = 1V$: $Mc_a$ is switched to $C_{min}$, $Mc_b$ is switched to $C_{max}$, and the output voltage is:}
    \begin{equation*}
      V_y = \frac{C_{max}}{C_{min} + C_{max}} * V_b > V_{HL} \approx logic \ 1, 
    \end{equation*}
   where $V_{HL}$ is the lower limit voltage for logic 1.
   \item{$Va = 1V, V_b = 0V$: with their connection polarities, $Mc_a$ is switched to $C_{max}$, $Mc_b$ is switched to $C_{min}$, and the output voltage is:}
    \begin{equation*}
      V_y = \frac{C_{max}}{C_{max} + C_{min}} * V_a > V_{HL} \approx logic \ 1 
    \end{equation*}
  \item{$Va = 1V, V_b = 1V$: the output voltage is:}
    \begin{equation*}
      V_y = \frac{Mc_a + Mc_b}{Mc_a + Mc_b} = 1V
    \end{equation*}
\end{itemize}

Here, the input combinations of $a$ and $b$ and the corresponding output values $y$ represent the truth table of an OR gate. Similarly, 3-input or 4-input AND and OR memcapacitive gates can be built. 

It is known \cite{Whitehead1912} that any logic expression can be described as a combination of AND, OR, and NOT functions. The NOT function generally requires an active element to complement its input signal. Since memcapacitive devices are passive, the NOT function cannot be implemented. As a consequence, we still need to rely on a traditional CMOS inverter to obtain a complete set of memcapacitive gates. 

\subsection{Memcapacitive Crossbar Classifier}
Crossbar architectures are attractive due to the regularity and the integration density. They have become more popular for memristive devices for these reasons \cite{Burger2014, Kataeva2015, Pouyan2015,Prezioso2015}. It has previously been shown that a general memcapacitive crossbar network can be built \cite{Strukov2013capacitive} and that such a crossbar network can perform a dot product \cite{GE2016Memcapacitive}.

\begin{figure}[!htb]
    \begin{center}
    {\includegraphics[width=\textwidth]{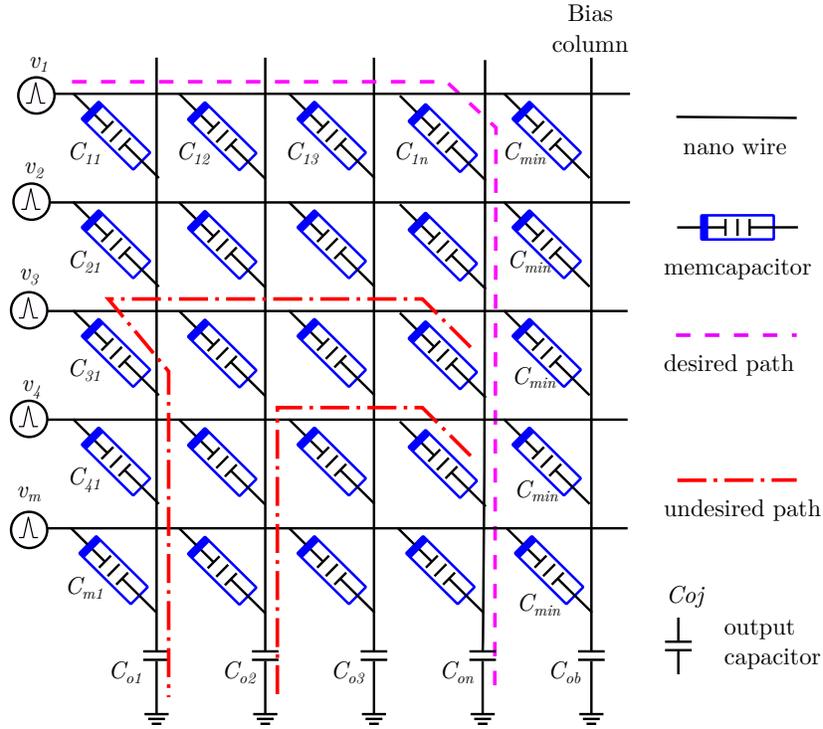}}
    \end{center}
    \caption{Memcapacitive crossbar network. The role of the termination output capacitors $C_{o1}, C_{o2}, C_{o3}, C_{o4}, C_{on}$ is to convert the total electric charge $q$ in each column $j$ to a corresponding voltage $V_{oj}$.}
    \label{fig:MemcapCrossbar}
\end{figure}

For our purpose, we propose the memcapacitive crossbar network as shown in Fig. \ref{fig:MemcapCrossbar}. This network functions as a classifier and can perform a dot product without the need of a processor and a memory as specified in \cite{GE2016Memcapacitive}. In this network, the memcapacitive devices are located at the nano-wire junctions. Each column has a termination capacitor $C_{oj}$ that converts the total charge in column $j$ to an equivalent voltage $V_{oj}$, which can then be measured. The crossbar also has a bias column. In our previous work on memristor crossbar architectures \cite{Woods2015}, we showed that a bias column is needed to compensate for currents in columns where all memristive devices are at $R_{max}$. $R_{max}$ represented a weight value of zero ($W = 0$), whereas $R_{min}$ represented a value of one ($W = 1$). Without a bias column, $R_{max}$ will still produce a small current in reality. The crossbar, which essentially computes a dot product, then results in an actual zero value when the bias column is used to compensate for the non-zero currents. This is essential for the training and testing of the crossbar classifier. 

We use the same approach for the memcapacitive crossbar classifier. From an electrical point of view, $C_{min}$ (the minimum capacitance of a memcapacitive device) at a column still allows a small charging current. Compensating for this current with the bias column ensures a zero dot product. In our memcapacitive crossbar network, all memcapacitive devices at the bias column were set to their minimum capacitance, which is equivalent to a zero weight.

\begin{figure}[!htb]
    \begin{center}
    {\includegraphics[width=0.4\textwidth]{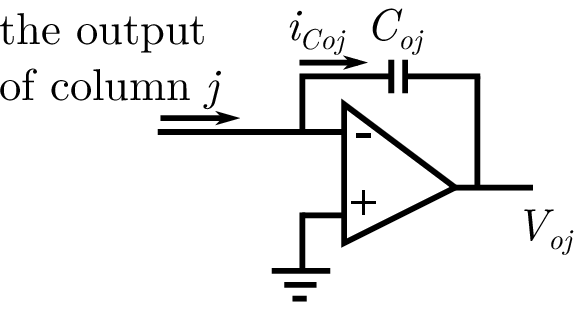}}
    \end{center}
    \caption{Virtual ground module. The OpAmp provides a virtual ground. Charge from output column $j$ is compensated by charging or discharging currents through the capacitor $C_{oj}$.}
    \label{fig:VirtGndModule}
\end{figure}

An inherent issue of any crossbar network is the effect of sneak-path currents. Several solutions have been proposed for memristive crossbar networks to overcome this problem: multistage readings \cite{Vontobel2009}, unfolded networks \cite{Manem2010}, complimentary reading algorithm \cite{Jung2012}, virtual ground \cite{Yakopcic2014}, or adapting three-terminal devices \cite{Zidan2013}. From Fig. \ref{fig:MemcapCrossbar}, the undesired paths allow additional charge from other columns to go to output capacitors $C_{o1}$ and $C_{o2}$, which then hold the total charge at only columns 1 and 2. 

For our memcapacitive crossbar, we propose a capacitive virtual ground module as shown in Fig. \ref{fig:VirtGndModule} at each output column. The 0V ground reference, provided by the OpAmp at each column $j$, eliminates all sneak-path currents. With the absence of sneak-path currents, voltage pulses at the input rows will charge the memcapacitive devices according to their internal capacitance states $\rho$. The total charge at a particular column $j$ is then accumulated and transferred to the output capacitor $C_{oj}$ in the virtual ground module. The total charge $Q_j$ at column $j$ is given by:

\begin{equation} \label{Eq:Cq}
    Q_j = \sum_{i=1}^{m} q_{i,j} - q_{bias},
\end{equation}
where $q_{i,j}$ is the electric charge stored in a memcapacitive device at the connective junction $(i,j)$ and $q_{bias}$ is the total electric charge of the bias column.

The subtractive term $q_{bias}$ ensures that the total charge $Q_j$ is zero when all memcapacitive devices at column $j$ are at their minimum capacitance. Expanding and simplifying Eq. \ref{Eq:Cq}, the output voltage $V_{oj}$ at output column $j$ becomes:
\begin{align}
    Q_j &= \sum_{i=1}^{m} V_iC_{i,j} - \sum_{i=1}^{m}V_iC_{min} \nonumber \\
    \left(-V_{oj}\right)C_{oj} &= \sum_{i=1}^{m} V_iC_{i,j} - \sum_{i=1}^{m}V_iC_{min} \nonumber \\
    -V_{oj} &= \sum_{i=1}^{m} V_i \left(\frac{C_{i,j} - C_{min}}{C_{oj}}\right), \label{Eq:VCout}
\end{align}
where $C_{i,j}$ is the capacitance of a memcapacitive device at junction $(i,j)$, bounded by the interval $\left[C_{min}, C_{max}\right]$.

Eq. \ref{Eq:VCout} shows that the output voltage at column $j$ is proportional to the device capacitance $C_{i,j}$ and $C_{min}$, the input voltage $V_i$, and the output capacitance $C_{oj}$. The output voltage $V_{oj}$ is independent of charge $Q_j$, the total charge of all memcapacitive devices at column $j$. As a result, our memcapacitive crossbar does not suffer the large effect of charge leakage as reported in \cite{Zidan2015compensated} for a MOS-gated memristor array.

\section{Results}
\subsection{Mem-devices in Logic Applications}
We used a pulse width $t_w$ and an amplitude $v_p$ to represent logic 1. To verify the logic gates, pulses were generated from the signal sources to simulate all the input states of a $n$-input gate. In addition, we measured the average power consumption of the memcapacitive gates and compared the values with equivalent memristive as well as CMOS gates.

According to \cite{Klimo2011memristors}, a valid output voltage of a memristive gate depends significantly on the changing states (switching from $R_{ON}$ to $R_{OFF}$ or vice versa) of the device and a high ratio of $R_{OFF}$ and $R_{ON}$. This changing state is linked to two physical factors of a memristive device, which vary from device to device: threshold voltage $v_{th}$ and switching time $t_s$. An applied pulse has to be sufficiently large ($v_p > 2v_{th}$) and long ($t_w > t_s$) so that the memristive devices can change their internal states and produce the correct outputs. Table \ref{tab:SwitchingTime} lists the switching times of all mem-devices we used here.


\begin{table}[!htb]
 \begin{center}
 \begin{adjustbox}{max width=\textwidth}
 \begin{tabularx}{\columnwidth}{|> {\em}l|X|c|r|c|c|}
 \hline
 \multirow{3}{*}{Model} & \multirow{2}{*}{Dev} & \multicolumn{2}{c|}{Pulse} & \multicolumn{2}{c|}{Switching Time of $\rho$}   \\
       \cline{3-6}
                       & & Ampl  & \multicolumn{1}{c|}{Width} & \multirow{2}{*}{$min \rightarrow max$} & \multirow{2}{*}{$max \rightarrow min$}\\
                       & Type & $v_p$ &\multicolumn{1}{c|}{$t_w$} &  & \\
 \hline
 Biolek \cite{Biolek2013} & $C$ & 2.4  & $2.0\mu s$ & $0.79\mu s$  & $0.80\mu s$  \\
 Mohamed\cite{Mohamed2015} & $C$ & 2.4  & $3.0s$     & $1.15s$      & $0.53s$      \\
 Chang \cite{Chang2013}              & $R$ & 2.4  & $7.0ps$    & $5.12ps$     & $2.23ps$     \\
 Oblea \cite{Oblea2010}  & $R$ & 2.4  & $500\mu s$ & $200.31\mu s$& $450.62\mu s$\\
 Sheridan \cite{Sheridan2011}        & $R$ & 4.85 & $60\mu s$  & $34.51ns$    & $23.63ns$    \\
 \hline
 \end{tabularx}
 \end{adjustbox}
 \end{center}
\caption{Switching times of mem-devices. $C$ stands for a memcapacitive and $R$ for a memristive device. $\rho$ is the internal state of a device. Each device was tested with a single pulse of amplitude $v_p$ and width $t_w$. The switching time was determined by measuring the change of its internal state from 1\% to 98\% ($min \rightarrow max$) or from 98\% to 1\% ($max \rightarrow min$) of its initial value.\\
}
\label{tab:SwitchingTime}
\end{table}

We selected three memristive models in Table \ref{tab:SwitchingTime} for their stability and their high $R_{OFF}/R_{ON}$ ratio. The \textit{Oblea} device had the lowest switching time of $450.62\mu s$ (at the exception of the \textit{Mohamed} device). We therefore used logic pulses of $500\mu s$ for all simulated mem-device gates and $3s$ pulses for the \textit{Mohamed} memcapacitive gates. Note that the switching time of the \textit{Mohamed} memcapacitive gate was so long because of the very slow convergence of the device's internal state from 1\% to 98\% of $Rho_{max}$ once it passed the 90\% point. The original \textit{Mohamed} memcapacitive model was developed for an input signal of 1V at 28.75MHz. We modified the model constants to accommodate a low frequency pulse signal. We targeted that time because we intended to use the memcapacitive device as a biologically plausible artificial synapse \cite{Rutherford2012, Song2012}. After a complete cycle, reset pulses were applied to reset the output of a gate before a new cycle began. 

\begin{figure}[!htb]
    \begin{center}
    {\includegraphics[width=\textwidth]{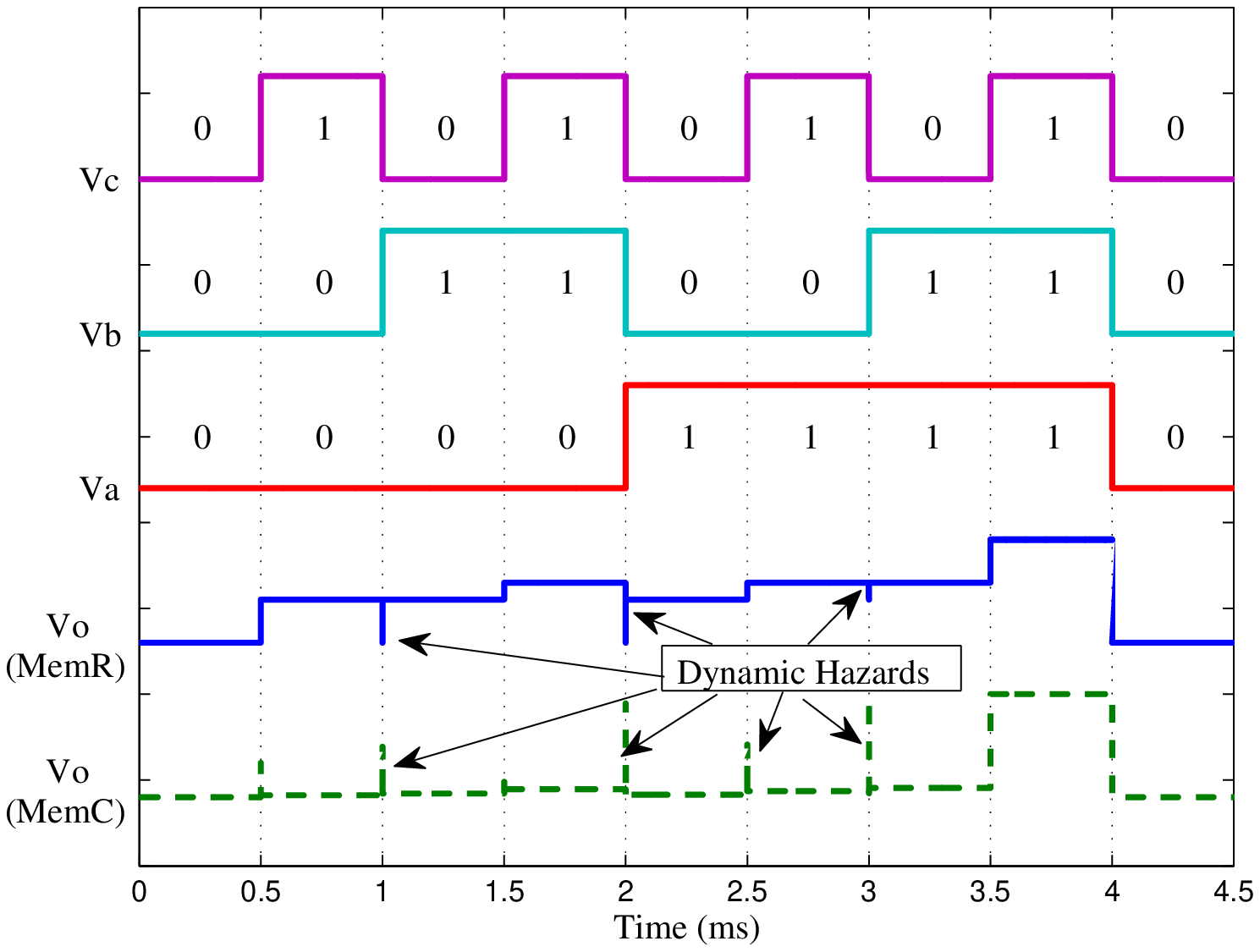}}
    \end{center}
    \caption{Timing diagrams of the 3-input AND gates (\textit{Sheridan} memristive gate and \textit{Biolek} memcapacitive gate). Each pulse was $500\mu s$ long. The bit values (0's and 1's) were added to show the input combinations of $V_a, V_b$, and $V_c$. $V_o(MemR)$ is the output of the memristive gate and $V_o(MemC)$ is the output of the memcapacitive gate.}
    \label{fig:3InputAND}
\end{figure}

Since a mem-device inverter cannot be built, we used CMOS inverters to build NAND, NOR, and XOR gates. For the full adder mem-device circuits, we utilized the mem-CMOS hybrid design of Cho \textit{et al.} \cite{Cho2015}. Their results showed that multilayer memristor-MOS circuits can implement any basic logic gate, such as AND, OR, NAND, NOR, and XOR. 

Fig. \ref{fig:3InputAND} shows the timing diagram of the 3-input mem-device AND gates. The bit values (0's and 1's) were added to show all input combinations. $V_o(MemR)$ and $V_o(MemC)$ show the outputs for the memristor and the memcapacitor gate respectively. Similar to memristive gates, our memcapacitive gates also showed dynamic hazards, a common phenomenon for memristive gates \cite{Kvatinsky2012}. Dynamic hazards occurred when the mem-devices switched their internal state $\rho$ (from $\rho_{min}\rightarrow \rho_{max}$ and vice versa). Within these transition times, the output logic was undefined.

\begin{figure}[!htb]
    \begin{center}
    {\includegraphics[width=\textwidth]{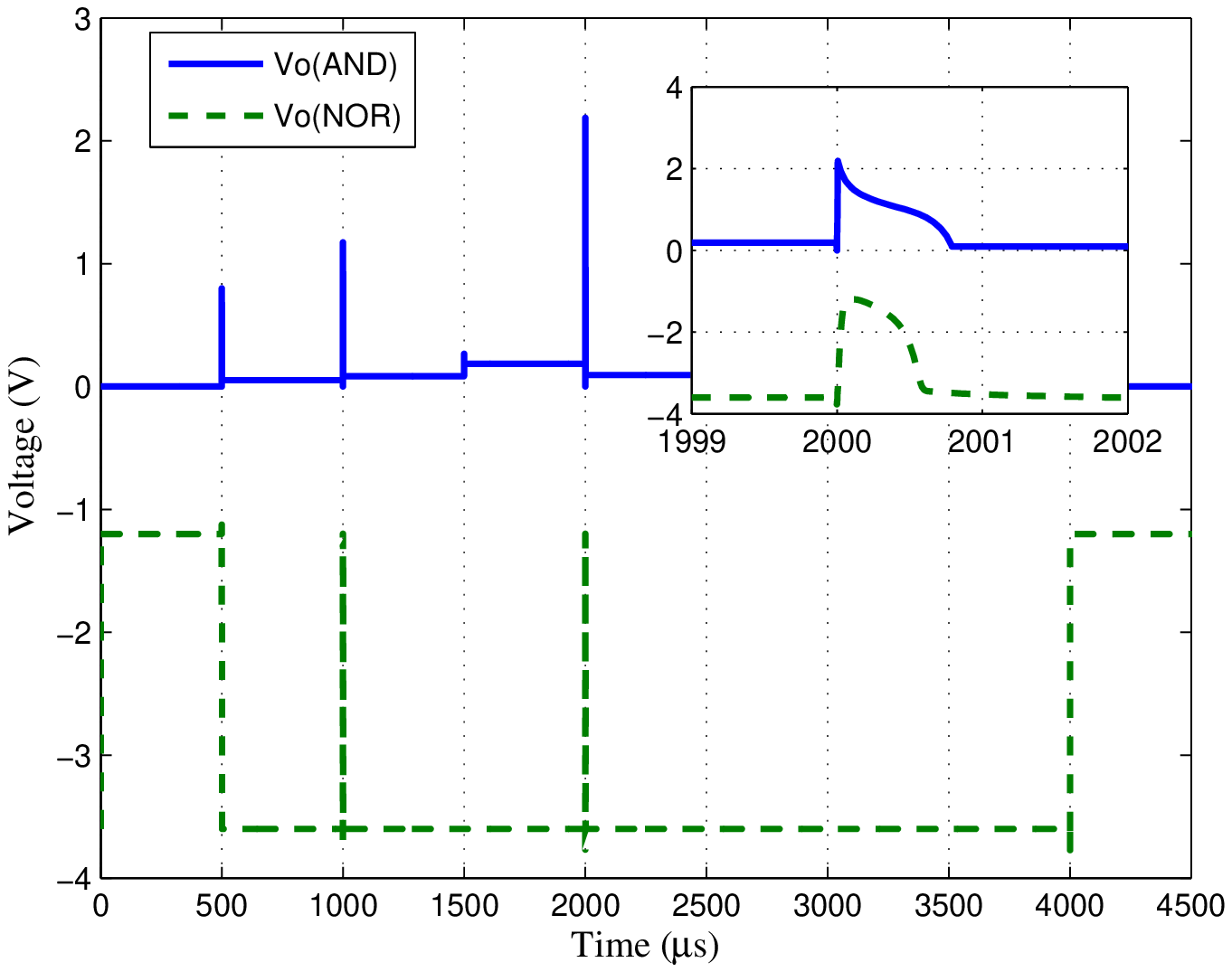}}
    \end{center}
    \caption{Dynamic logic hazards of 3-input memcapacitive AND(-) and NOR(--) gates using the \textit{Biolek} model. The inset shows the output values in the interval [$1999\mu s, 2002\mu s$]. With a $500\mu s$ pulse, the spike width of a logic hazard was about $0.8\mu s$ for the memcapacitive AND gate and about $0.6\mu s$ for the memcapacitive NOR gate.}
    \label{fig:HazardSpike}
\end{figure}

The inset in Fig. \ref{fig:HazardSpike} shows a dynamic hazard of the memcapacitive AND and NOR gates in the interval $[1999\mu s, 2002\mu s]$. The spike width estimates were $0.8\mu s$ and about $0.6\mu s$ for AND and NOR gates respectively. With a pulse width of $500\mu s$, dynamic hazards can be potentially avoided by adding a time delay before reading the outputs. Another approach to remove dynamic hazards is to add buffers or inverters along the signal paths to restore the logic signals \cite{Kvatinsky2012}.

\begin{figure}[!htb]
    \begin{center}
     {\includegraphics[width=\textwidth]{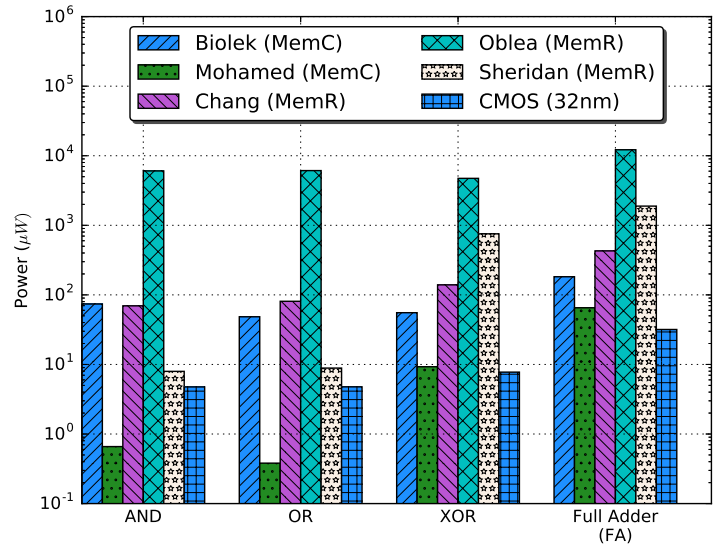}}
     \end{center}
    \caption{Power consumption of the mem-device logic and CMOS gates. The \textit{Biolek} and \textit{Mohamed} gates are memcapacitive gates, the rest are memristive gates. The overall winners for the mem-device gates were the \textit{Mohamed} memcapacitive gates.}
    \label{fig:GateBarCharts}
\end{figure}

Fig. \ref{fig:GateBarCharts} shows the power consumptions for the mem-device gates as well as for 32nm CMOS gates. The power consumption was determined with SPICE by using RMS measurements of voltages and currents over a complete cycle of $2^n$ pulses (where $n$ is  the number of inputs of a gate). For AND and OR gates, the \textit{Sheridan} memristive gates consumed less power than the \textit{Biolek} memcapacitive gates. The overall winners were the \textit{Mohamed} memcapacitive gates. Furthermore, the \textit{Sheridan} memristive gates required a 4.8V pulse amplitude whereas the memcapacitive gates only needed 2.4V. A lower voltage can be an advantage for interfacing with CMOS devices that are operated in a low-power mode \cite{Das2014}. 

The memcapacitive circuits outperformed the memristive circuits in terms of power consumption for XOR and FA. Compared to CMOS gates, the \textit{Mohamed} memcapacitive AND and OR gates used less power. The memcapacitive XOR and full adder circuits, however, used more power than CMOS circuits due the CMOS inverters that are needed to implement NOT functions. In fact, the power consumptions of the CMOS inverters for the mem-device XOR and full adder circuits contributed about $95\%$ to the total power consumptions. 

\begin{table} [!htb]
\begin{center}
\begin{adjustbox}{max width=\textwidth}
\begin{filecontents*}{GatesRes.csv}
Gate,BiolekC4,Mohamed,Chang,Oblea,Sheridan,CMOS
2-input AND,53.1478,0.288925,35.486,4174.03,4.54666,3.58626
3-input AND,84.8712,0.780228,68.4431,6479.05,8.37162,4.75659
4-input AND,83.8331,0.90975,104.715,7492.31,10.9431,5.91193
% Gate,BiolekC4,Mohamed,Chang,Oblea,Sheridan,CMOS
2-input OR,47.9053,0.288904,35.486,4174.03,4.25202,3.59923
3-input OR,51.4109,0.386708,83.5016,6658.72,8.29032,4.77142
4-input OR,45.9856,0.46845,123.672,7516.97,14.0635,5.92512
% Gate,BiolekC4,Mohamed,Chang,Oblea,Sheridan,CMOS
2-input NAND,54.456,2.19045,57.8326,4184.75,297.054,2.80034
3-input NAND,85.5395,2.98594,91.0065,6489.45,363.685,3.91076
4-input NAND,84.7346,3.35248,132.97,7501.04,380.415,5.00439
% Gate,BiolekC4,Mohamed,Chang,Oblea,Sheridan,CMOS
2-input NOR,6.08801,1.73186,63.6967,4180.42,296.741,2.63307
3-input NOR,224.961,2.29963,117.31,6667.63,366.31,3.73078
4-input NOR,10.266,2.3015,158.124,7525.44,392.405,4.8402
% Gate,BiolekC4,Mohamed,Chang,Oblea,Sheridan,CMOS
2-input XOR,57.0868,8.36312,102.446,4198.28,542.086,5.62276
3-input XOR,53.8057,10.1649,176.131,5263.6,965.532,9.88489
% Gate,BiolekC4,Mohamed,Chang,Oblea,Sheridan,CMOS
1-input FA,151.338,45.2674,273.169,10993.1,1164.14,22.1184
2-input FA,213.097,85.1835,584.062,13367.4,2611.59,41.5025
\end{filecontents*}
\pgfplotstabletypeset[
    fixed zerofill,
    precision=2,
    col sep=comma,
    comment chars=\%,
    assign column name/.style={/pgfplots/table/column name={\em{#1}}},
    every head row/.style={
       before row={\hline},
       after row={
        & {MemC} & {MemC} & {MemR} & {MemR} & {MemR} & 32nm\\
& {($\mu W$)}  & {($\mu W$)}   & {($\mu W$)}  & {($\mu W$)}  & {($\mu W$)} & {($\mu W$)} \\
        \hline},
    },
    column type/.add={|}{},
    columns/Gate/.style={string type},
    every last column/.style={column type/.add={}{|}},
    every row no 3/.style={before row=\hline},
    every row no 6/.style={before row=\hline},
    every row no 9/.style={before row=\hline},
    every row no 12/.style={before row=\hline},
    every row no 14/.style={before row=\hline},
    every last row/.style={after row=\hline},
    ] {GatesRes.csv}

\end{adjustbox}
\end{center}
\caption{Summary of mem-device gates' power consumption. Logic pulses of $v_p$ and $t_w$ were applied to simulate all input combinations. CMOS inverters were used for the mem-based NAND, NOR, XOR, and FAs.\\
}
\label{tab:GatesRes}
\end{table}

Table \ref{tab:GatesRes} summarizes the results of our simulations. CMOS inverters were used for the mem-based NAND, NOR, XOR, and FAs. We compared 
the average power consumptions of the memristive gates (\textit{Chang}, \textit{Oblea}, and \textit{Sheridan}) and the CMOS gates with those of the \textit{Mohamed} memcapacitive gates (the overall winners) for power saving factors. The results of the power saving factors are shown in Table \ref{tab:SavingFactors}.


\begin{table}[ht!]
\begin{center}
\begin{tabularx}{0.75\columnwidth}{
    |X
    |S [table-format= 3.1, round-mode = places, round-precision = 1]
    |S [table-format= 5.1, round-mode = places, round-precision = 1]
    |S [table-format= 3.1, round-mode = places, round-precision = 1]
    |S [table-format= 2.1, round-mode = places, round-precision = 1]
    |}
\hline
\multirow{2}{*}{Gate} & {\textit{Chang}} & {\textit{Oblea}} & {\textit{Sheridan}} & {\textit{CMOS}} \\
 & {MemR} & {MemR} & {MemR} & {32nm} \\
\hline
AND  & 105.43 &  9169.42 & 12.06  & 7.2\\
OR   &  212.1 & 16039.09 & 23.26  & 12.5 \\
NAND &  33.04 &  2131.03 & 122.07 & 1.37 \\
NOR  &  53.55 &  2901.24 & 166.66 & 1.77 \\
XOR  &  15.04 &   510.68 & 81.37  & 0.84 \\
FA   &   6.57 &   186.74 & 28.94  & 0.49 \\
\hline
\end{tabularx}
\end{center}
\caption{Power saving factors when comparing the average power consumptions of the memristive gates and the CMOS gates with those of the \textit{Mohamed} memcapacitive gates.}
\label{tab:SavingFactors}
\end{table}


These results show that memcapacitive gates are a promising option for implementing low-power digital logic circuits.

\subsection{Mem-devices in Crossbar Classifiers}
A classifier often functions as an output layer, for example in deep learning networks for image processing and pattern recognition. In a pattern recognition application, a classifier is trained in a supervised way, in which expected outputs are provided along with the input images. Once the training process is completed, the classifier is tested with a different set of image data for how well it can recognize similar patterns. We trained and tested our mem-device crossbar classifiers with two typical datasets: MNIST \cite{Domhan2015} and CIFAR-10 \cite{Lee2015}. The MNIST dataset contains handwritten digits of size $28\times28$. This dataset has 60,000 training and 10,000 testing images. The CIFAR-10 dataset is a collection of 60,000 color images of size of $32\times32$, which is divided into 50,000 training and 10,000 testing images. There are 10 different classes of objects. 

\begin{figure}[!htb]
    \begin{center}
    {\includegraphics[width=\textwidth]{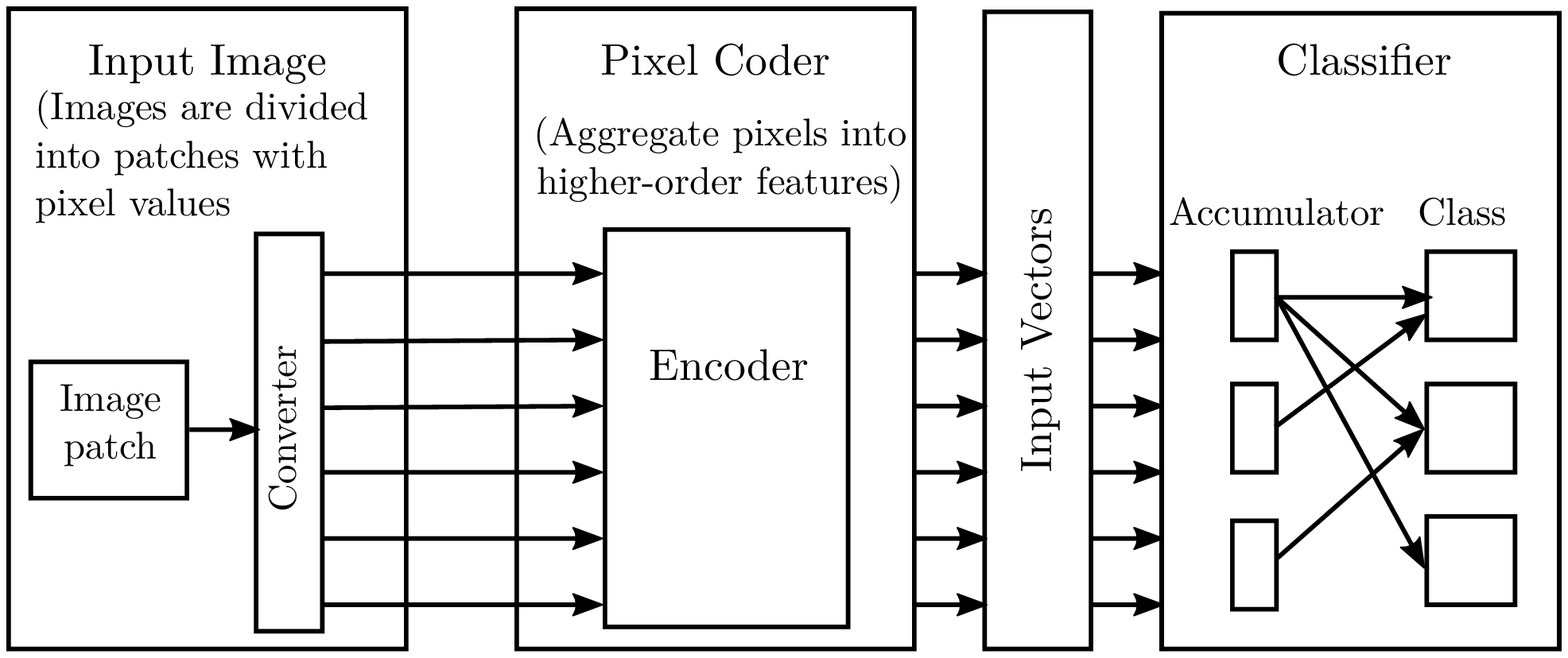}}
    \end{center}
    \caption{An example of an artificial network that performs image recognition.}
    \label{fig:ImageProcessing}
\end{figure}

Fig. \ref{fig:ImageProcessing} shows an example of a network performing pattern recognition that we employed for training and testing our memcapacitive classifiers. In this network, training and testing images are divided into smaller patches of pixel values. The converter then converts image pixels into input values for the coder. The coder encodes the pixel inputs, aggregates these inputs into higher-order features of input images, and produces input vectors for training and testing the classifiers.

We first trained the mem-device crossbar classifiers and then tested the classification performance. We also calculated the average power consumption per image for both the training and testing phases.

\begin{figure}[!htb]
    \begin{center}
    {\includegraphics[width=\textwidth]{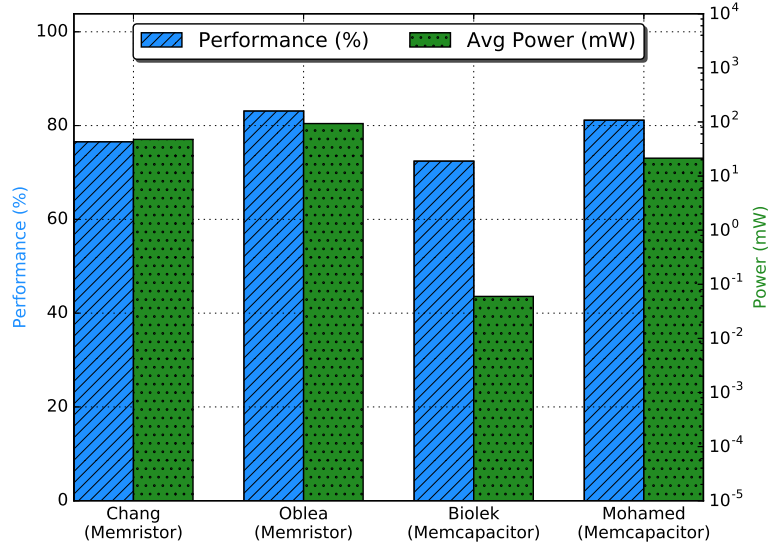}}
    \end{center}
    \caption{Classification performance and power consumption of the mem-device classifiers for the MNIST dataset. The power measurements do not include the power consumption of the virtual ground modules.}
    \label{fig:BarMNIST}
\end{figure}

The training stage of a classifier, particularly a mem-device crossbar classifier, was composed of two phases: the inference phase and the update phase. In the inference phase, the outputs of the classifier were collected with applied training data while the internal states of mem-devices remained unchanged. We normalized the input vectors to ensure that the input voltages were less than the threshold voltages for the mem-devices and that the mem-devices did not change their internal states during the inference phase. In the update phase each mem-device was updated individually based on the feedback from a supervised learner. The supervised learner used gradient descent with back-propagation to determine how to update each mem-device with a $250\mu s$ pulse. The $250\mu s$ pulse is specific to the \textit{Chang} memristive device and we used it for all classifiers. Once the classifiers were trained, they were tested with test images for clarifications. Both the training and testing stages were performed in Python. The average power was determined as the average power consumed by all mem-devices during the inference phase, the update phase, and the testing phase.

\begin{figure}[!htb]
    \begin{center}
    {\includegraphics[width=\textwidth]{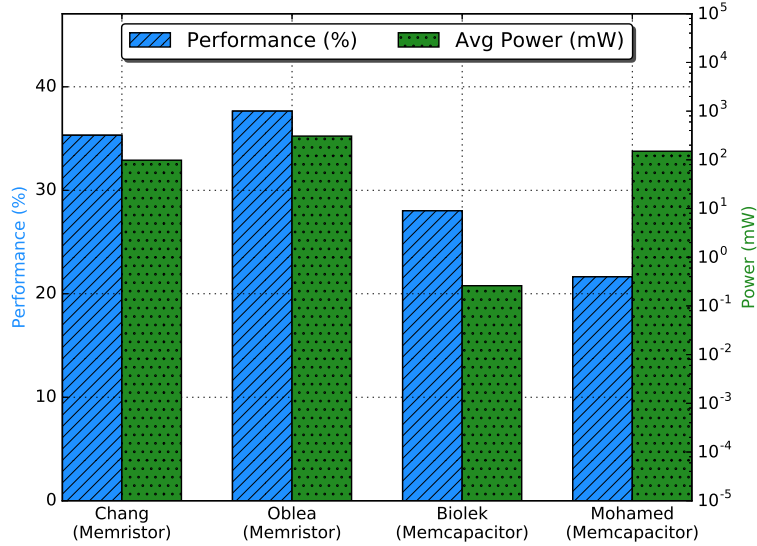}}
    \end{center}
    \caption{Classification performance and power consumption of the mem-device classifiers for the CIFAR-10 dataset. The power measurements do not include the power consumption of the virtual ground modules.}
    \label{fig:BarCIFAR10}
\end{figure}

Fig. \ref{fig:BarMNIST} shows the simulation results of the mem-device classifiers for the MNIST dataset. The mem-device crossbar had a size of $1568\times10$. The classifier size was determined by the input image vectors. These vectors were generated by the sparse and independent local network (SAILnet) algorithm for the MNIST dataset, which has $14\times14$ patches with over-completeness of 2. SAILnet utilized an improved model to represent a more realistic response of a mammalian visual cortex \cite{Zylberberg2011}. The results show that our memcapacitive classifiers performed similarly compared to the memristive classifiers while they consumed less power per image on average. As one can see, the \textit{Biolek} memcapacitive classifier has the lowest power consumption of all models. 

Fig. \ref{fig:BarCIFAR10} compares the simulation results of the mem-device classifiers for the CIFAR-10 dataset. In order to maintain a reasonable size of our mem-device classifiers (such as $4508\times10$), the color images were converted to gray scale images for training and testing. Furthermore, a whitening process was applied to the input images in order to reduce the highly correlated adjacent pixels, which showed to improve both the training time and performance \cite{Hariharan2012}. The length of each input image vector determined the size of the classifiers. For CIFAR-10, the SAILnet algorithm generated the input vectors of $16\times16$ patches and an over-completeness of 2.

As one can see from Fig. \ref{fig:BarCIFAR10}, the memcapacitive classifiers did not reach the performance of memristive classifiers, but they consumed less power. The performance of memcapacitive classifiers correlated directly with the setting parameters (the learning rate $\alpha$, the update pulse width $t_w$, the update pulse amplitude $v_w$, and the offset voltage $v_{\textit{offset}}$) during the training phase. These parameters were chosen based on experiments.

We suspect that the memcapacitive classifiers do not reach the performance of the memristive classifiers for the following reason: since we do not have positive and negative weights, $v_{\textit{offset}}$ is used to so that weight $W$ is set between $C_{min}$ and $C_{max}$ after a training phase. If $v_{\textit{offset}}$ is low, most weights are bound to $C_{min}$. If $v_{\textit{offset}}$ is high, most weights are set to $C_{max}$. For the MNIST dataset, the inputs are very sparse and we can, therefore, find a reasonably good value of $v_{\textit{offset}}$ experimentally. On the other hand, the inputs of the CIFAR-10 dataset are not sparse enough. As a result, a small change in $v_{\textit{offset}}$ causes the entire weight matrix to be shifted to either $C_{min}$ or $C_{max}$. The memristive classifiers seem to be less sensitive to the $v_{\textit{offset}}$ value, and, therefore, perform better.

\begin{table} [!htb]
\begin{center}
\begin{adjustbox}{max width=\textwidth}
\begin{tabularx}{\columnwidth}{
    |> {\em}l
    |> {\centering}X
    |S [table-format= 2.2, round-mode = places, round-precision = 2]
    |S [table-format= 2.3, round-mode = places, round-precision = 3]
    |S [table-format= 2.2, round-mode = places, round-precision = 2]
    |S [table-format= 3.3, round-mode = places, round-precision = 3]
    |}
\hline
\multirow{4}{*}{Model} & \multirow{2}{*}{Device}  & \multicolumn{4}{c|} {Dataset}\\
    \cline{3-6}
                       &      & \multicolumn{2}{c|}{MNIST}  & \multicolumn{2}{c|} {CIFAR-10} \\
    \cline{3-6}
                       & \multirow{2}{*}{Type} & {Perf.} & {Crossbar} & {Perf.} &  {Crossbar}\\
                       &      & {(\%)}  &  {$(mW$)}  & {(\%)}     &     {$(mW)$}   \\
\hline
Chang \cite{Chang2013}       & $MemR$ & 76.52 & 47.60694166 &   35.32 & 98.35331742 \\
Oblea \cite{Oblea2010}       & $MemR$ & 83.08 & 93.40705699 &  37.65  & 307.3198393 \\
Biolek \cite{Biolek2013}     & $MemC$ & 72.4  & 0.05966735132 & 28.02 &  0.2600728912 \\
Mohamed\cite{Mohamed2015}    & $MemC$ & 81.13 & 21.40676618 &  21.64  & 150.3011448 \\
\hline
\end{tabularx}
\end{adjustbox}
\end{center}
\caption{Summary of the classification performance and power consumption. The power measurements do not include the power consumption of the virtual ground modules. The power measurements were averaged over each image for both the training and testing phases.}
\label{tab:Performance}
\end{table}

Table \ref{tab:Performance} shows a summary of the simulation results. Using the average power consumption of the \textit{Biolek} memcapacitive classifier as a reference, we compared its results with those of the \textit{Chang} and \textit{Oblea} classifiers. For the MNIST dataset, the \textit{Biolek} classifier could achieve equal classification performance and save power by factors of $797\times$ and $1565\times$ respectively. For the CIFAR-10 dataset, the \textit{Biolek} classifier saved power by factors of $378\times$ and $1181\times$. 
\section{Discussion}
As it was shown in Table \ref{tab:SwitchingTime}, the \textit{Oblea} device has the slowest settling time with the exception of the \textit{Mohamed} device. As a result, we used $500\mu s$ pulses to test all mem-device logic gates. Operating mem-device logic gates with $500\mu s$ pulses is quite slow compared to CMOS logic gates. However, the \textit{Biolek} memcapacitive logic gates with a smaller switching time are capable to operate with $2\mu s$ pulses.

Both memristive and memcapacitive gates suffered the effect of dynamic hazards. Dynamic hazards occurred when the mem-devices of a gate switched their internal states. Therefore, a delay time was required before the gate's output could be read. This delay time is similar to the setup time in a CMOS gate, although the CMOS setup time is much smaller. Recent studies have shown that new memristive devices can switch their internal states much faster (in the range of ns and ps) \cite{Kim2015charge, Choi2016high}. A faster switching time would imply less dynamic hazards. 

The \textit{Mohamed} memcapacitive XOR and the full adder circuits did not outperform the CMOS circuits in terms of power consumption. However, about $95\%$ of the power consumption was due to the CMOS inverters and transistors that are required for the gates in addition to the mem-devices. 

The performance of the memcapacitive classifiers depends on how the memcapacitive devices are updated. The process involves setting four parameters: the learning rate $\alpha$, the update pulse width $t_w$, the update pulse amplitude $v_w$, and the offset voltage $v_{\textit{offset}}$. These parameters were based on experiments. A systematic exploration of the parameter space is beyond the scope of this paper. We expect that the classification performance can be further increased with better parameters. 

Moreover, virtual ground modules played an essential role in alleviating the effect of sneak-path currents within the crossbar networks. We have left out the power figures for these modules because they are highly technology-dependent.  

\section{Conclusion}
Our work has shown that low-power memcapacitive logic circuits can be implemented. The memcapacitive gates consumed about $7\times$ less power compared to memristive logic gates. The lack of a mem-inverter makes the possible logical basis incomplete. The inverter operation, by its nature, requires an active element to reverse its input signal, which cannot be realized by passive mem-devices. Used for classifiers, memcapacitive devices were shown to reduce the power consumption by a factor of $1,500\times$ for MNIST and a factor of $1,000\times$ for CIFAR-10. For the classifier, we relied on virtual ground modules, which remove the effects of sneak-path currents, but consume significant power. Finding other options to eliminate sneak-path currents without the need of virtual ground modules could further lower the power consumption. 

\section{Acknowledgments}
This work was supported by the Defense Advanced Research Projects Agency (DARPA) under award \# HR0011-13-2-0015. The views expressed are those of the author(s) and do not reflect the official policy or position of the Department of Defense or the U.S. Government. Approved for Public Release, Distribution Unlimited.

The authors also thank Jens B\"{u}rger and Walt Woods for the helpful discussions.
\appendix

\end{document}